# Answering Hilbert's 1st Problem

Charles Sauerbier

> *Dogma gives a charter to mistake, but the very breath of science is a contest with mistake, and must keep the conscience alive.* ~ George Eliot

## Abstract

Hilbert's first problem is of importance in relation to work being done in computational systems. It is the question of equipollence of natural and real numbers. By construction equipollence is established for real numbers in open interval (0, 1) and natural numbers and, from such to all real numbers. Construction stands in contradiction of the generally accepted diagonal argument of Cantor. Mathematics being irrefutable, in absence rejection of all theory of mathematics and logic, the problem exists in acceptance; that itself arises of more fundamental a problem in science generally. The problem within Hilbert's problem is of Schopenhauer's, et al, "will and representation" born.

## Background

Hilbert in lecture, as recorded in [1], produced at the turn of the last century a list of problems in mathematics. The first problem in his list is titled "*Cantor's Problem of the Cardinal Number of the Continuum*". By all available references this remains an open problem a century after being posited. The question is itself central to many open problems in computational systems theory. In answering Hilbert's challenge so distant in time one opens new problems, and upends some of what has come to be accepted.

The question of whether the set of real numbers is denumerable predates both Cantor and Hilbert. Smorynski in [2] presents Cantor's "*Pairing Function*", that is an element in proof of work presented here. [3] is a general reference for author's substantive recollection of concepts that is of particular note for breadth of content of relevance to presented work.

## Problem within the Problem

> *"A given opinion, as held by several individuals, even when of the most congenial views, is as distinct as are their faces."* ~ St. John Henry Cardinal Newman

Such is at the root of the argument in proof of what is written. Homage here need be paid to Hofstadter's classic [4]; in which pages one finds perhaps the most eloquent exposition of what here is only briefly touched, after Louis Carol, in the vein of the insightful comment of Cardinal

Newman and others. It is hopefully with sufficiency necessary to move beyond the inherent fallacy of reasoning giving rise to it.

A *Gedankenexperiment* if you will: You come upon a literal "." point in the emptiness of space. From it, as far into the distance as distant can be you observe this tiny machine tirelessly and endlessly extending a string of what by all appearance is the digit "3" on your left from the point; only to observe an identical tiny machine is similarly doing at the end of identical string extending to your right.

Someone then appears, of nowhere particular, to inquire of what purpose these tiny machines, and of what is their intent. Your first thought of what those two machines are doing?

Therein rests foundation of Hilbert's consternation. Our two tiny machines are in fact doing nothing more than what is literally observed: extending from a single point in space two strings of what in our ideation represent a digit – namely "3". All else of any response arises of an individual's mental notions and processes – of *opinions held*, as it be.

In [5] Arthur Schopenhauer comments at first line what is, irrespective opinions held, reality: "The world is my representation"[1]. He gives in that work worthy discussion of the metaphysical problem within Hilbert's 1st problem. While in [6] and in [7] one is introduced to how science goes so awry that Carrol and Hofstadter should have sufficient material of which to write such beguiling text, so entertaining of the mind.

If you were told our two tiny machines, having heard Einstein's conjecture the universe was curved in manner of a giant sphere, were intent on encircling it with means of conveyance for which the "rail" was composed of what did to all observers appear to be the digit "3"? What then might your will represent of your experience? Thus only to have a second figure appear, of nowhere particular, to explain the two tiny machines are writing the results in both directions from their origin of $\sum_{n=0}^{\infty} 3 \times 10^{n}$, inviting you to draw nearer that point in space; where upon you become aware that point is an ever expanding sphere in which you suddenly find yourself enveloped, and come to observe that from this point extend an infinity of what you perceive as strings of digits, in what do appear of random orderings of digits extended and branched endlessly in all directions by tiny machines that tirelessly append "*digits*" at the end of each sequence and spontaneously appear, of nowhere particular, to append nine branches from each digit in each apparent string – "*My God, it's full of Stars*" ("2001: A Space Odyssey").

You have experienced the enumeration of real numbers? Of the fractional part? Of the natural numbers? Reality arises of your ideation as to meaning of what is observed. That is the problem most essential here. Mathematics stands irrefutable on its own dry foundations.

---

[1] Which in its original text potentially infers much more in meaning through consensual ideation of words writ.

## Introduction and Generalities of Import

What is presented is premised on basic, commonly known, generally accepted principles for which proofs are readily available in basic theory text and taught in basic theory courses at undergraduate level. But a few for reader's reference of influence in part on work herein: [8] [9] [10].

Fundamental to the "*Continuum Problem*" are notional concepts as such exist of our individual and consensual ideation and representation thereof. From perspective of metaphysics, what in our mind we conceptualize as "*numbers*" do not exist beyond that discrete realm of our thoughts, where such notional concepts (ideations) exist: All else is representation.

What it means to be effectively computable was reduced to its essence and put forth by Muḥammad ibn Mūsā al-Khwārizmī's in "*The Compendious Book on Calculation by Completion and Balancing*", and the many subsequent works of others: *A set of symbols and a set of rules for their manipulation.* Herein is given the manipulation of symbols so as to establish a bijection from real numbers into the natural numbers.

Proof of equipollence of real numbers and natural numbers is accomplished by constructively establishing a bijection between generally accepted representations of natural numbers and real numbers, with the primary category defining distinction of said numbers being the fractional part of real numbers.

From the set of symbols $S$ = { 0, 1, 2, 3, 4, 5, 6, 7, 8, 9 } (digits), together with adjunct set { . } (period) a set of strings is recursively enumerated. That set being simply the *Kleene Closure* of $S$, prepended with singular element of { . }; from such then is taken that set that is the fractional parts of real numbers, as proper subset.

Let $D$ represent that set of strings on $S \cup \{ . \}$ such that all strings $d$ in $D$ are thus conforming:

$d$ = <period> + <digit_sequence>
<digit_sequence> = <digit> | <digit> + <digit_sequence>
<digit> = 0 | 1 | 2 | 3 | 4 |5 | 6 | 7 | 8 | 9

By convention and generally accepted theory $D$ must contain all possible sequences that can occur as fractional part of any real number that is not an integer. It happens that $D$, as defined, will contain sequences that are, by convention and generally accepted theory, not permissible as the fractional part of a real number. Such is of no consequence in fact to necessary bijection, as any subset of a countably-many set is itself. By elimination of impermissible sequences from $D$ set $F$ is obtained; where $F$ is that set consistent with the consensual human ideation of real numbers.

## Construction of Set D

Let $S$ = { 0, 1, 2, 3, 4, 5, 6, 7, 8, 9 } be a set of symbols. Let '.' (period) be a root symbol. Let $T$ be a set of trees, $t_i$, where index $i$ = 0, 1, 2, 3, …, such that: (1) $t_0$ has only a single vertex (both root and leaf) labeled with symbol '.', and; (2) $t_i$ is obtained from $t_{i-1}$ by appending one vertex for each symbol in $S$ to each leaf vertex in $t_{i-1}$. Such trees be that which our tireless tiny machines recursively enumerated in endless labor. Each element $t_i$, for $i$ greater than 0, in $T$ have two important properties: (1) every inner vertex has degree 10, and; (2) no vertex descendent of any (parent) vertex has the same label as another descendent of that vertex. Property (2) equivalently stated: *No two siblings of a common parent are labeled with same element of S*.

Each tree $t_i$ in $T$ has finite and countable number of leaves. Each sequence in each tree of labels from root to any leaf is distinct, by construction. Each set of sequences extracted from each tree is itself countable by common and generally accepted theory; in addition to the cardinality of each set of such sequences being computable by means within the commonly known and generally accepted theory. All sequences defined by each $t_i$, taken collectively as a single set, constitute the *Kleene Closure* on $S$. The set $T$, indexed by positive integers, is a denumerable set of infinite cardinality; in which, as elements, there exists the representation of sets, individually, of finite and infinite cardinality. Each set itself denumerable in consequence of having been recursively enumerated by construction.

It is well known and generally accepted that the union of recursively enumerable sets is itself denumerable. The set of sequences obtained by union of all the sequences produced of each $t_i$ in $T$, is by consequence countable, hence denumerable. It is of $T$, by its construction, that $D$, by its definition, can be obtained in representation; $D$ thus being denumerable.

It is here that ideation puts before each reader what for many presents as an impassable impediment to acceptance: What does it mean that a string is of "*potentially*" infinite length. All readers might share a consensual ideation of "*infinite*", in context of strings as extending without end. So what then makes the sequence (string) produced by $\sum_{n=0}^{\infty} 3 \times 10^{n}$ of a different "infinity" that that produced by $\sum_{n=1}^{\infty} 3 \times 10^{-n}$? Aside from our ideation in regards the former extending to the left of a period and the latter extending to the right in the windmills of our mind with which we tilt? Our argument here, of which each reader must in their own mind accept or reject, is that both are composed of "countably many" symbols and in that fact differ naught in concept of infinity. Infinity is in both cases equivalent. Our construction here of $D$ should make such mathematically and logically irrefutable. Only because we stated $D$ was intended to recursively enumerate all strings that could serve as fractional part of a real number does reader accept such. If told $D$ recursively enumerated the set of string from which all natural numbers can be taken as subset how could what is "*infinite*" change, beyond how each reader conceives infinite to be in their mind. Mathematics is irrefutable as to what infinite is, and the horse sufficiently ridden dead.

## Of Set D take Set F as Proper Subset

Set *D* is denumerable by construction through recursive enumeration. It is, similarly, as a set constructed, containing the *Kleene Closure* on *S*, and; is composed of countably-many elements, and; is of "*infinite*" cardinality. *D* contains *F*. *F* being that set of all possible sequences that can exist as fractional part of any real number. From the mathematical realities of *D* set *F* as a proper subset of *D* is of as infinite cardinality, in any sense, as *D*.

*D* also contains sequences not permissible as fractional part in our consensual ideation of real numbers. *D*, for instance, contains that set of strings containing only the 0 (zero) digit symbol and period in initial position. *D* also contains as sequences those that contain no digit save 0 beyond some ordinal position. While none of this denies what is mathematically irrefutable at this point as to factional part of real numbers being equipollent to natural numbers, means exists to reduce *D* to *F*.

Let *Z* be the set of sequences in *D* which beyond some ordinal position only the 0 (zero) symbol exists. *Z* is therefore that set of sequences respective the 0 (zero) symbol that are not allowed as fractional part of any real number. Set *Z*, as a proper subset of a countable set, is countable.

Let *F* then be obtained by union of $\varepsilon$ (empty sequence) and relative complement of *Z* in *D* *(D \ Z)*. *F* thus contains all elements of *D* except those elements also in *Z*. *F* then by construction contains that set of sequences that are allowable fractional part of any real number. *F* remains denumerable as proper subset of a denumerable set.

## Of Empirical Objections What Said?

By similar exclusion the reader can further reduce the set *F* thus obtained to remove any strings that offend their ideation of what is a fractional part of a real number to obtain that set of strings agreeable to them. Irrespective what is removed the resulting set remains countable. It thus retains, where not reduced to a finite set, a bijection with the set of natural numbers; absent reader's objection that the *Kleene Closure* on any finite set of symbols produces a "*finite set*" of strings of "*finite length*". To such reader one might suggest a review of category theory. By such the <u>*set*</u> produced by the closure is of "*countably-many*" (infinite) cardinality. It contains as strings all possible strings, to include those strings of symbols that are of "*countably-many*" (infinite) ordinal positions in length. By consensual representation of what is denoted by *countably-many* the *Kleene Closure* is a set of infinite cardinality, in which there exist sets; of which some such sets are of infinite cardinality in consequence of being, as well, of infinite ordinal length as representation of what is by will ideation as strings or sequences of symbols.

## Of Thus Enumeration of Real Numbers

Cantor's basis for diagonal argument was premised on a restriction on *F* as representation of the set of fractional parts of real numbers in open interval (0, 1). Such set is by construction given and shown, in contradiction of Cantor's conclusion to the contrary, to be denumerable – in a bijection placed with the set of natural numbers as an infinite subset of an infinite set for which such bijection is proven by recursive enumeration of elements.

Using Cantor's "*Pairing Function*" on their Cartesian product the sets integers and *F*, as obtained above, one obtains a bijection of real numbers with natural numbers; as the Cartesian product contains all possible representations of our ideation of real numbers. Perhaps Schopenhauer's true intent was "[reality] is my representation"?

Cantor's fault of reason exists in Gödel's subsequent proof as to completeness where taken in context of representation: A set of higher dimensionality cannot be represented as a set in lesser dimensionality. Cantor's list being in 2 dimension constructed could never contain a set of dimensionality greater than 2. Such is in part the essence of Hofstadter's recursion of sets.

## Elle est la raison

There exists then reason giving rise to necessity? In the abstract consider existence of a single universal set of dimension unobservable; that by observation appears to contain a finite set of things. Each element of that set being a single set similarly composed. And so recursively for as long as one might continue to examine elements of any element. But in observation one notes that individually each set is not of necessity composed of "symbols" from one common set, as were each set in case of real numbers, except where one at each encounter of a symbol not previously accounted adds such to the set of symbols considered.

Stepping back to look again from without one has a recursive construction, from what might be an infinite set of symbols, a "structure" of what might be infinite dimensions within a finite constraint. So giving cause to question if in the limit is what is potentially within the structure more numerable than the real numbers? By construct the answer must be no. Mirroring the real numbers, up to the point of real numbers being in each recursion constrained to a defined finite set of symbols, it is obvious that the structure is a "tree" growing in potentially unbounded depth and breadth, and yet at no point, even in the limit, not enumerable.

The structure is an abstraction of any system, and abstraction of any system's progression in time. It fundamentally gives rise to Bell's curse of dimensionality in application of *Dynamic Programming* to systems; at the same time giving rise to potential means to overcome Bell's curse, even if within some limit.

## Consequence

Recursive enumeration of the set of infinite cardinality that contains all strings, of both finite and infinite ordinal number of symbols composed, such that each string therein uniquely comprises a representation of a single fractional part of our collective notion of such as part of our representation of such conceptual notion denoted as "*real numbers*" is established by construction. The construction presented makes uses only of well-known and widely accepted mathematical and computational theory. The bijection of real numbers with natural numbers thus proven is, irrefutably in context of mathematics and logic.

## Conclusion

The set of real numbers and set of natural numbers are equipollent, as proven beyond refutation in mathematics and logic alone, and; thus give in answer of Hilbert's 1st problem: No.

The humorous element is that by reason is derived an answer that by common knowledge and generally accepted theory is an *effective computation* computable by a Turing Machine, but one that would never halt so as to render the answer to the operator; save to God alone could of such the answer be obtained, at the end of time. But then the question did arise of nowhere particular save of will and representation of the one, and the dissimilarity of the one ideation in the minds of the many.

## Contributions & Acknowledgment

*Fiorella Marincich-Sauerbier* – Contribution to key and critical conceptual points not rising to commonly accepted standard for authorship; though invaluable in work presented and efforts to reduce thought to (hopefully) cogent representation in a human comprehensible language.